\documentclass[aps,twocolumn,prb]{revtex4}
\usepackage{amsmath}
\usepackage{bm}
\usepackage{color}
\usepackage{epsfig}

\newcommand{\B}{\mbox{\tiny B}}

\newcommand{\D}{\mbox{\tiny D}}
\newcommand{\ind}{{\sf n}}
\newcommand{\rhonup}{\rho_{\sf n}^{{ }_{\{\!+\!\}}}}
\newcommand{\rhondown}{\rho_{\sf n}^{{ }_{\{\!-\!\}}}}
\newcommand{\rhonpm}{\rho_{\sf n}^{{ }_{\{\!\pm\!\}}}}

\newcommand{\la}{\langle}
\newcommand{\ra}{\rangle}

\newcommand{\Sec}[1]{Sec.\,\ref{#1}}

\newcommand{\be}{\begin{equation}}
\newcommand{\ee}{\end{equation}}
\newcommand{\bsube}{\begin{subequations}}
\newcommand{\esube}{\end{subequations}}
\newcommand{\Eq}[1]{Eq.\,(\ref{#1})}
\newcommand{\Eqs}[1]{Eqs.\,(\ref{#1})}
\newcommand{\Fig}[1]{Fig.\,\ref{#1}}
\newcommand{\Figs}[1]{Figs.\,\ref{#1}}

\newcommand{\NP}{[N\!+\!1/N]}
\newcommand{\NN}{[N/N]}
\newcommand{\NM}{[N\!-\!1/N]}

\begin{document}

\title{
 Optimized hierarchical equations of motion for Drude dissipation
}

\author{Jin-Jin Ding,$^1$
  Jian Xu,$^{2}$   Jie Hu,$^{2}$
  Rui-Xue Xu,$^{1,\ast}$}

\author{YiJing Yan$^{1,2,}$}\email{rxxu@ustc.edu.cn; yyan@ust.hk}
\affiliation{$^1$Hefei National Laboratory for Physical Sciences at
the Microscale, University of Science and Technology of China,
Hefei, Anhui 230026, China and \\
$^2$Department of Chemistry, Hong Kong University of
Science and Technology, Kowloon, Hong Kong}

\date{Submitted to J.\ Chem.\ Phys.\ on 22 June 2011}

\begin{abstract}

The hierarchical equations of motion theory
for Drude dissipation is optimized,
with a convenient convergence criterion proposed
in advance of numerical propagations.
The theoretical construction is on basis of a
Pad\'{e} spectrum decomposition that has been qualified to be the best
sum-over-poles scheme for quantum distribution function.
The resulting hierarchical dynamics under the
{\em apriori} convergence criterion are exemplified with a benchmark
spin-boson system, and also the transient absorption and
two-dimensional spectroscopy of a model exciton dimer system.

\end{abstract}
\maketitle

\section{Introduction}
\label{thintro}

 Quantum dissipation plays a pivotal
role in many chemical, physical, and biological
problems.\cite{Nit06,Wei08,Kle09,Yan05187}
Recent experiments on excitation energy transfer in photosynthesis
systems\cite{Eng07782,Col10644} show clearly the breakdown
of conventional Markovian and perturbative
quantum dissipation theories.\cite{Ish09234110,Ish09234111}
The protein environment is nano-structured, with the time
scale of modulation comparable
to that of the excitation energy transfer.
Also the protein-pigment coupling strength is
about that between pigments.
Apparently, one needs
a nonperturbative, non-Markovian, and also
numerically implementable quantum dissipation theory.
The demand is the same on the study of heating generation in quantum transport
through mesoscopic systems\cite{Seg023915,Gal081056,Pop10147}
and protection of qubits with the aid of photonic crystal
environment.\cite{Lod04654,Sap101352,Wol10141108}

 This work focuses on the hierarchical equations of motion (HEOM) approach.%
\cite{Tan906676,Ish053131,Tan06082001,Yan04216,Xu05041103,Xu07031107}
We request the best HEOM construction, exemplified with Drude dissipation, together
with a convenient criterion to estimate
its convergency before propagations for general systems
at any finite temperature.
HEOM approach was originally proposed in 1989
by Tanimura and Kubo
for semiclassical dissipation.\cite{Tan89101}
Formally exact HEOM formalism\cite{Tan906676,Ish053131,%
 Tan06082001,Yan04216,Xu05041103,Xu07031107} %
for Gaussian dissipation in general, including its second
quantization,\cite{Jin08234703} is now well established, with the
aid of proper environment spectrum decomposition schemes. As a numerically
efficient alternative to path integral influence
functional,\cite{Fey63118,Wei08,Kle09} HEOM has been applied to such
as electron transfer,%
 \cite{Shi09164518,Xu09214111,Tia10114112,Tia10114112note,Tan10214502}
nonlinear optical spectroscopy,%
 \cite{Ish06084501,Ish079269,Tan091270,Che10024505,Che11194508,Zhu115678}
and transient quantum transport.%
\cite{Zhe08184112,Zhe08093016,Zhe09164708}

To have an explicit HEOM construction one should exploit
certain basis set spanning over the stochastic bath space.
This is equivalent
to the choice of certain sum-over-pole
(SOP) scheme that decomposes
individual bath correlation function into
its multiple-timescale memory spectrum components.
The conventional scheme is the Matsubara expansion
of quantum distribution
function (i.e., the Bose/Fermi function) involved in the bath correlation functions.
\cite{Wei08,Kle09,Yan05187}
However, Matsubara expansion is notorious for slow convergence.
 We have recently proposed three Pad\'{e} spectrum
decomposition (PSD) schemes\cite{Hu10101106,Hu11jcp}
be the candidates for the best SOP method. Mathematically,
these three PSDs
of Bose/Fermi function exploit the
$\NM$, $\NN$, and $\NP$ Pad\'{e} approximants, respectively.
The resulting HEOM dynamics have been demonstrated in context
of transient quantum transport through a double quantum
dots system and population transfer in a spin-boson system.\cite{Hu11jcp}

  Three closely related issues arise in
the choice of statistical environment basis set (or SOP scheme)
for efficient HEOM construction.
The first one is to identify
the {\em best} basis set for spanning
over the stochastic bath space.
The second one concerns the {\em minimum} basis set.
It requests to have not only the smallest
number of decomposition terms, but also
a priori accuracy control criterion
on the resulting HEOM dynamics
for any given system under bath influence at finite temperature.
The third issue is about
the possibility of at least partial
inclusion of the off-basis-set
residue effect on the HEOM dynamics.

 In this work, we construct an optimized
HEOM theory for Drude dissipation.
We identity that $\NN$ PSD serves as the
best and the minimum Drude dissipation basis set.
It leads naturally to an off--basis--set
white noise residue (WNR) term, together with
a simple accuracy control criterion on the resulting HEOM dynamics.
The present paper is a generalization
of Ref.~\onlinecite{Xu09214111} and Ref.~\onlinecite{Tia10114112},
where $N=0$ and $N=1$ cases were analyzed, respectively.
We summarize the PSD-based HEOM formalism
with the Drude bath in \Sec{ththeo},
followed by proposing the accuracy control criterion.
Numerical demonstrations are carried out in \Sec{thnum}.
Included there are a benchmark
spin-boson dynamics, studied before by Thoss, Wang,
and Miller,\cite{Tho012991}
and the nonlinear optical signals
of a model dimer system.
Finally we conclude the paper in \Sec{thsum}.

\section{Formalism}
\label{ththeo}

\subsection{Optimal hierarchy for Drude dissipation}
\label{ththeoA}

 In this subsection, we exploit the $\NN$ PSD scheme\cite{Hu11jcp}
to construct HEOM for Drude dissipation cases.
HEOM has the following generic form,\cite{Xu07031107}
\be\label{dotrhon}
 \dot\rho_{\ind}(t)=
  -[i{\cal L}(t)+\gamma_{\ind}+\delta{\cal R}_\ind]\rho_{\ind}(t)
  +\rhondown(t) +  \rhonup(t).
\ee
It describes how an $n^{\rm th}$-tier ADO $\rho_{\sf n}$
depends on its associated $(n\pm 1)^{\rm th}$-tier
ADOs in $\rhonpm$.
The ADO's index $\ind$ is in general a collection
of indices; i.e., $\ind\equiv \{n_1,\cdots,n_K\}$, with $n_k\geq 0$ for
bosonic bath. Here $n_1+\cdots+n_K=n$ for
$\rho_{\sf n}= \rho_{n_1,\cdots,n_K}$ being called
an $n^{\rm th}$-tier ADO.
The reduced system density operator
$\rho(t)\equiv {\rm tr}_{\B}\rho_{\rm total}(t)
\equiv \rho_{\sf 0}(t)$ is just
the zeroth-tier ADO.
In \Eq{dotrhon}, the reduced system Liouvillian ${\cal L}(t)\,\cdot\, \equiv
[H(t),\,\cdot\,]$ can be time dependent, e.g., in the
presence of driving fields.
Throughout of this paper, we set  $\hbar=1$
and $\beta=1/(k_BT)$, with
$k_B$ being the Boltzmann constant and $T$ the temperature.

 The specific HEOM construction, including the ADO
labeling index $\ind=\{n_1,\cdots,n_K\}$,
depends on the way of decomposing bath correlation
function into its memory spectrum components.
For clarity, let the system-bath interaction
be $H'(t)=-{\hat Q}{\hat F}_{\B}(t)$, with $\hat Q$
and $\hat F_{\B}(t)$ being operators in
the reduced system and the stochastic bath subspaces,
respectively.
The influence of bath on system is
completely determined by the correlation
function $C(t) \equiv \la \hat F_{\B}(t)\hat F_{\B}(0)\ra_{\B}$.
It is in turn related to the bath spectral density $J(\omega)$
via the fluctuation-dissipation theorem (FDT):\cite{Wei08,Kle09,Yan05187}
\be\label{fdt0}
 C(t) = \frac{1}{\pi} \int_{-\infty}^{\infty}\!\!d\omega
  \frac{e^{-i\omega t}J(\omega)}{1-e^{-\beta\omega}}.
\ee
To have a HEOM construction,\cite{Xu07031107}
we expand $C(t)$ in a finite exponential series,
on the basis of certain SOP scheme, together with
the Cauchy residue theorem of contour integration.

In this work we focus on the Drude model,
\be\label{Jdrude}
  J(\omega)
 =\frac{2\lambda\gamma \omega}{\omega^2+\gamma^2}.
\ee
It has only one pole, $z=-i\gamma\equiv -i\gamma_{\D}$, in the lower-half plane.
The exponential series of bath correlation function assumes then
\be\label{fdt}
 C(t) = \sum_{k={\rm D},1}^{N}\!\! c_k e^{-\gamma_kt}+\delta C_N(t).
\ee
In general the off-basis-set residue $\delta C_N(t)\neq 0$, as one
exploits only finite $N$ poles for Bose function.
The conventional scheme is the Matsubara expansion which is however
notorious for slow convergence.\cite{Wei08,Kle09}
For Drude dissipation it has been suggested\cite{Tia10114112} that
$\NN$ PSD be the best SOP of Bose function.
It is accurate up to ${\cal O}(x^{4N+1})$ in the order of $x=\beta\omega$
and reads \cite{Hu11jcp}
\be\label{pade}
 \frac{1}{1-e^{-x}}
\approx  f^{\NN}(x) =
 \frac{1}{x}+\frac{1}{2}
 +  \sum_{k=1}^{N} \frac{2\eta_kx}{x^2+\xi^2_k}+ R_N x  ,
\ee
with
\be\label{RN}
  R_N = \frac{1}{4(N+1)(2N+3)}\,.
\ee
The PSD poles and coefficients, $\{\xi_k,\eta_k;k=1,\cdots,N\}$,
are all positive and can be evaluated with high precision via
the eigenvalues of real symmetric triangle matrices.\cite{Hu11jcp}

 The corresponding exponential series expansion of \Eq{fdt} can now be obtained
via the standard contour integration technique.
We obtain (setting $\gamma_k\equiv \xi_k/\beta$)
\be \label{drude}
\begin{split}
  &c_{\D}=-2i\lambda\gamma f^{\NN}(\beta z)\big|_{z=-i\gamma} ,
\\
 &c_k = \frac{2\eta_k}{i\beta}J(z)\big|_{z=-i\gamma_k};
     \ k=1,\cdots,N ,
\end{split}
\ee
and
\be\label{delCN}
 \delta C_N(t) \approx 4\lambda\beta\gamma R_N\delta(t)\equiv 2\Delta_N\delta(t).
\ee
Note that $\{c_k; k=1,\cdots,N\}$ are all real. On the other hand,
$c_{\D}$ associating with the Drude exponent
$\gamma_{\D} =\gamma$ is complex and is
evaluated by using the $\NN$ Bose function $f^{\NN}(x)$,
as given by
the last expression of \Eq{pade}.

 The SOP scheme resembles a statistical environment basis set for the HEOM construction.
It dictates not only the exponential expansion
of $C(t)$ as \Eq{fdt}, but also HEOM.
The ADO reads now $\rho_{\sf n}\equiv \rho_{n_{\D},n_1,\cdots,n_N}$.
The only approximation involved
is the WNR treatment of the off-basis-set $\delta C_N(t)$ by \Eq{delCN}.
It results in the WNR of $\delta{\cal R}_{\ind}$ in \Eq{dotrhon}
without further approximation:\cite{Ish053131,Xu07031107}
\be\label{calR_WNR}
 \delta{\cal R}_{\ind}\rho_{\ind}
 = \Delta_N[\hat Q,[\hat Q,\rho_{\ind}]].
\ee
The damping parameter in \Eq{dotrhon} collects
all relevant exponents:\cite{Xu07031107}
\be\label{gam_ind}
  \gamma_{\ind}= \sum_{k={\rm D},1}^{N} n_k \gamma_k.
\ee
The tier-down and tier-up terms are \cite{Xu07031107,Shi09084105}
\be\label{updown}
\begin{split}
    \rhonup& =-i\sum_{k={\rm D},1}^{N}
   \sqrt{(n_{k}+1)|c_k|}\,\bigl[\hat Q, \rho_{{\ind}_{k}^+}\bigr]\,,
\\
  \rhondown &=-i\sum_{k={\rm D},1}^{N}
    \sqrt{\frac{n_k}{|c_k|}}\,
     \bigl(c_k \hat Q \rho_{{\ind}_{k}^-}
       -c^\ast_k\rho_{{\ind}_{k}^-}\hat Q\bigr)\, ,
\end{split}
\ee
with $\rho_{{\ind}_{k}^{\pm}}$ being
the associated $(n\pm 1)^{\rm th}$-tier ADO, respectively.
The labeling index ${\ind}_{k}^{\pm}$
differs from $\ind$ only by changing
the specified $n_k$ to
$n_k\pm 1$.
All ADOs here are dimensionless and scaled properly
for the efficient HEOM propagator
with the recently developed on-the-fly filtering algorithm.\cite{Shi09084105}
Numerically it also automatically truncates
the level of hierarchy.\cite{Shi09084105}

\subsection{Accuracy control criterions}
\label{ththeoB}

 As mentioned earlier the only approximation
involved is the WNR treatment of the off-basis-set
$\delta C_N(t)$ by \Eq{delCN}.
Its validity dictates therefore
the accuracy of the resulted HEOM.
The exact residue $\delta C_N(t)$, which is a real and even function,
is defined via \Eq{fdt}, together with \Eq{drude} for the Drude model.
Its spectrum $\delta C_N(\omega)\equiv
 \tfrac{1}{2}\int dt\,e^{i\omega t}\delta C_N(t)$
is a symmetric and bell-shaped function,
being positive and monotonically decreasing in $\omega>0$
from $\delta C_N(\omega=0)=\Delta_N$
to $\delta C_N(\omega\rightarrow\infty)=0$,
where $\Delta_N=2\lambda\beta\gamma R_N$ was defined in \Eq{delCN}.
The half-width-at-half-maximum $\Gamma_N(\gamma)$ that
characterizes the inverse time scale of $\delta C_N(t)$
is determined via
$\delta C_N(\omega)\vert_{\omega=\Gamma_N(\gamma)}=\Delta_N/2$.
It is found that $\Gamma_N(\gamma)$ can be well approximated by (within
$0.5\%$ of relative error for $N\leq 16$ as tested)
\be\label{GamN}
 \Gamma_N(\gamma) \approx
  \frac{1}{\beta}\Big[r_N
    +\sqrt{(\beta\gamma)^2+0.34 r_N^2}\,
    \Big]\,,
\ee
where $r_N=1/(2R_N)=2(N+1)(2N+3)$ [cf.\ \Eq{RN}].

\begin{figure}
\includegraphics[width=0.95\columnwidth]{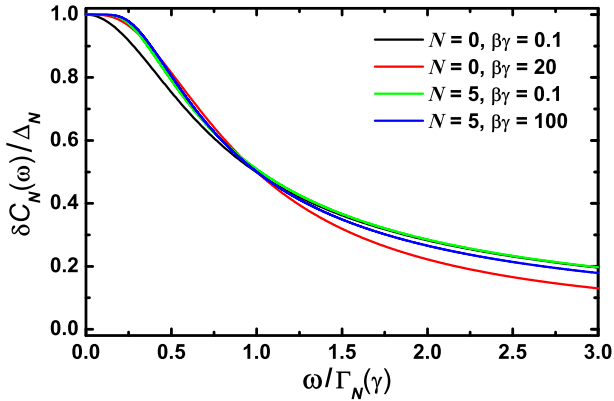}
\caption{The deviation spectrum function
 $\delta C_N(\omega)$ of Drude bath, plotted
 in terms of $\delta C_N(\omega)/\Delta_N$ versus $\omega/\Gamma_N(\gamma)$
 for some selected values of $\{N, \beta\gamma\}$,
 where $\Delta_N$ is given via \Eq{delCN} and
 $\Gamma_N(\gamma)$ is by the approximant of \Eq{GamN}.
}
\label{fig1}
\end{figure}

 Figure \ref{fig1} depicts the residue
spectrum $\delta C_N(\omega)$, plotted in terms of
$\delta C_N(\omega)/\Delta_N$ versus $\omega/\Gamma_N(\gamma)$
for some selected values of $\{N, \beta\gamma\}$.
Used here for the $x$-axis scaling
is the approximated $\Gamma_N(\gamma)$ value of \Eq{GamN}.
Thus this figure shows also the excellent quality of \Eq{GamN}.

\begin{figure}
\includegraphics[width=0.95\columnwidth]{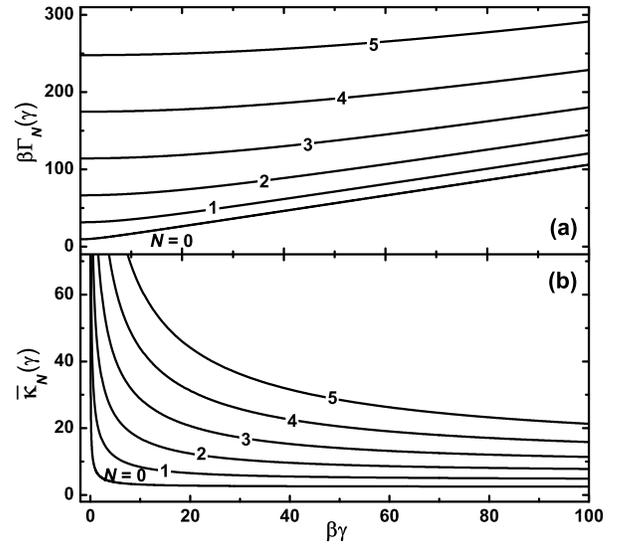}
\caption{$\beta\Gamma_N(\gamma)$ and
 $\bar\kappa_N(\gamma)\equiv
 (\beta\lambda)^{1/2}\kappa_N(\gamma,\lambda)$
 versus $\beta\gamma$.
}
\label{fig2}
\end{figure}

 To validate the $\delta$-function approximation
of $\delta C_N(t)$ [\Eq{delCN}],
we examine the Kubo's motional narrowing line shape parameter,\cite{Kub63174,Kub69101}
\be\label{kappaN}
 \kappa_N(\gamma,\lambda)  =\sqrt{\Gamma_N(\gamma)/\Delta_N}  =
 \sqrt{r_N\Gamma_N(\gamma)/(\beta\lambda\gamma)}\, .
\ee
The evaluated $\beta\Gamma_N(\gamma)$ and
$\bar\kappa_N(\gamma)\equiv
  (\beta\lambda)^{1/2}\kappa_N(\gamma,\lambda)$,
as functions of $\beta\gamma$,
are depicted in \Fig{fig2}(a)
and (b), respectively.

 The accuracy control criterion on the HEOM
in \Sec{ththeoA} comprises therefore the
conditions under which $\delta C_N(t)$ and its effect on the reduced
system dynamics can be treated as Markovian
white noise.
Apparently, it is valid when
$\Gamma_N(\gamma)\gg \Omega_s$  and
$\kappa_N(\gamma,\lambda)  \gg 1$,
with $\Omega_s$ denoting the characteristic frequency
of system.
We have found\cite{Xu09214111,Tia10114112} that the HEOM dynamics assumes numerically
accurate (if not exact) when
\be\label{crit5}
 \min\{\Gamma_N(\gamma)/\Omega_s, \kappa_N(\gamma,\lambda)\} \gtrsim 5,
\ee 
while semi-quantitative when
\be\label{crit2}
 2\lesssim
   \min\{\Gamma_N(\gamma)/\Omega_s, \kappa_N(\gamma,\lambda)\}
 \lesssim 5.
\ee
The above accuracy control criterions facilitate the choice of
minimum $N$ for the desired quality of HEOM dynamics;
see Refs.\ \onlinecite{Xu09214111} and
\onlinecite{Tia10114112} for the special
cases of $N=0$ and $1$, respectively.

\section{Numerical demonstrations}
\label{thnum}

\subsection{Spin-boson dynamics}
\label{thnumA}

 To demonstrate the efficiency of HEOM and also the proposed
accuracy control criterions, consider a benchmark spin-boson system,
studied before by Thoss, Wang,
and Miller via their numerically exact self-consistent hybrid approach.\cite{Tho012991}
The system Hamiltonian is
$H=\epsilon\sigma_z+V\sigma_x$, with the dissipative mode of $\hat Q=\sigma_z$,
subject to Drude dissipation.
Here $\sigma_x$ and $\sigma_z$ are Pauli matrices. The Rabi frequency of the
system is $\Omega_s=2\sqrt{\epsilon^2+V^2}$.
We choose the most challenging parameters used in Ref.\ \onlinecite{Tho012991},
i.e., the figure 8 there, with $\epsilon/V=1$, $\lambda/V=0.25$, $\gamma/V=5$,
and $\beta V=50$.

\begin{figure}
\includegraphics[width=0.95\columnwidth]{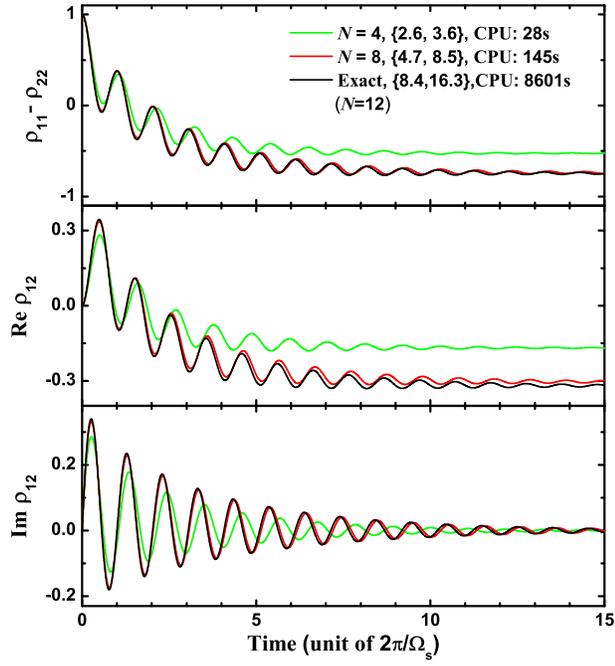}
\caption{Evolution of the reduced spin system density matrix elements,
with the same parameters as Fig.\ 8 of Ref.\ \onlinecite{Tho012991};
i.e., $\epsilon/V=1$, $\lambda/V=0.25$, $\gamma/V=5$, and $\beta V=50$.
The accuracy control parameters are
$\{\Gamma_N/\Omega_s,\kappa_N\} =\{2.6,3.6\}$ for $N=4$ and
$\{4.7,8.5\}$ for $N=8$,
as indicated. The converged dynamics are identical
to that of $N=10$ whose $\{\Gamma_N/\Omega_s,\kappa_N\}
=\{6.3,12.0\}$.
}
\label{fig3}
\end{figure}

Figure \ref{fig3} depicts the reduced system density matrix evolution,
as evaluated from the HEOM at each specified environment basis set size $N$.
It converges when $N\geq 10$ (by eye inspection).
The accuracy control parameters
$\{\Gamma_N/\Omega_s,\kappa_N\}$ are $\{2.6,3.6\}_{N=4}$,
$\{4.7, 8.5\}_{N=8}$, $\{6.3, 12.0\}_{N=10}$,
and  $\{8.4, 16.3\}_{N=12}$, respectively.
The CPU time listed in \Fig{fig3} is based on
a single Intel(R) Xeon(R) processor@3.00GHz.
The HEOM propagation uses the fourth-order
Runge-Kutta method with time-step of $0.001/V$, together with
the on-the-fly
filtering algorithm \cite{Shi09084105} with error tolerance of $5\times10^{-7}$.
Evidently the numerically accurate criterion (\ref{crit5}) and the
semi-quantitative one (\ref{crit2}) are both verified.
Moreover, the $\NN$ PSD-based HEOM is found to be
the best among all schemes we tested.

\subsection{Nonlinear spectroscopy of a model exciton dimer system}
\label{thnumB}

 We now examine the accuracy control over the $\NN$ PSD-based HEOM in
evaluating nonlinear spectroscopic signals.
We will verify again the proposed accuracy
control criterions in \Eqs{crit5} and (\ref{crit2}), respectively.

\begin{figure}
\includegraphics[width=0.95\columnwidth]{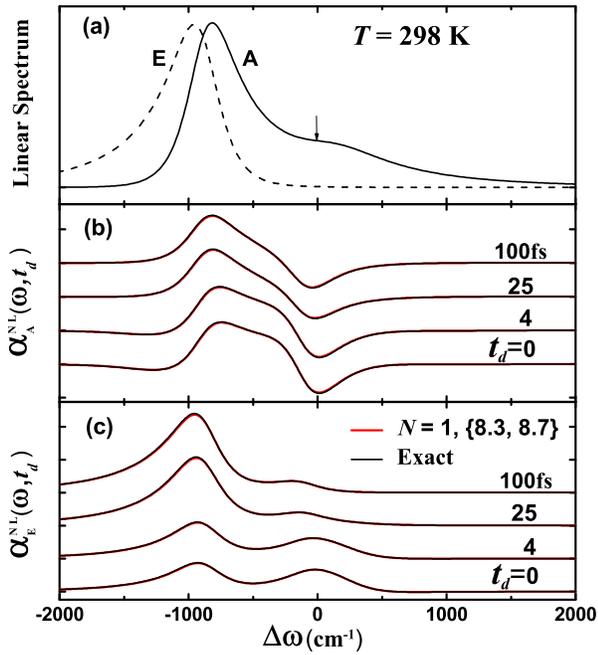}
\caption{Dispersed transient absorption coefficient signals
of the model
dimer system (see text):
(a) Linear absorption (A)
and emission (E) signals; (b) Nonlinear absorptive
$\alpha^{\rm NL}_{\rm A}(\omega,t_d)$ component;
(c) Nonlinear emissive $\alpha^{\rm NL}_{\rm E}(\omega,t_d)$
component. The pump field is a transform-limited 50\,fs-pulse
of finite intensity (see text), centered at
$\omega=\epsilon$; i.e.,
$\Delta\omega\equiv \omega-\epsilon=0$,
as indicated by the arrow in panel (a).
At $T=298$\,K, $N=1$ is in the numerically accurate range, with
$\{\Gamma_N/\Omega_s,\kappa_N\}=\{8.3,8.7\}$ as indicated.
}
\label{fig4}
\end{figure}

\begin{figure}
\includegraphics[width=0.95\columnwidth]{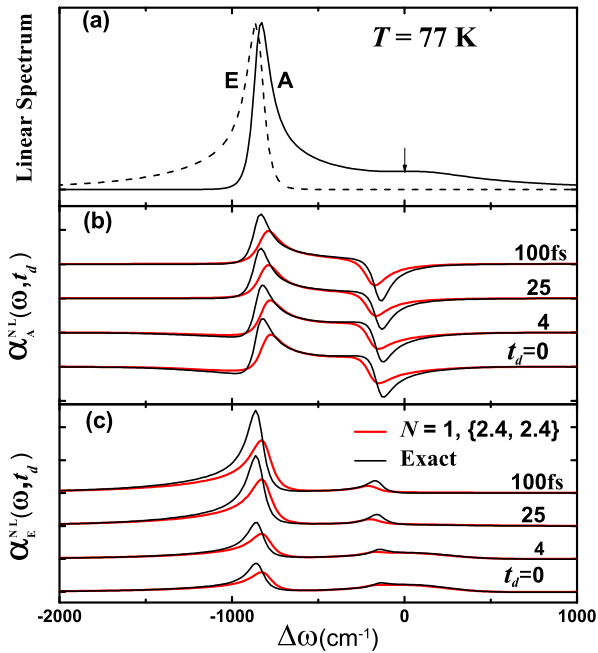}
\caption{Same as \Fig{fig4}, but at $T=77$\,K, where
$N=1$ is in the semiquantitative range, with
$\{\Gamma_N/\Omega_s,\kappa_N\}=\{2.4,2.4\}$, as indicated.}
\label{fig5}
\end{figure}

We first re-evaluate the figures 6 and 7 of Ref.\ \onlinecite{Tia10114112},
due to a careless mistake in the excitation field configuration used
there.\cite{Tia10114112note}
The correct signals (in rotating wave approximation) are now depicted
in \Figs{fig4} and \ref{fig5}, respectively.
Demonstrated are different components of the
dispersed transient absorption coefficient
$\alpha(\omega,t_d)$ of a model exciton dimer
system at two representing temperatures.
Here, $t_d$ denotes the probe
delay time with respect to the pump pulse.
Denote also $\Delta\omega=\omega-\epsilon$,
where $\epsilon=\epsilon_1=\epsilon_2$ is the
on-site excitonic energy.
The dimer system parameters\cite{Tia10114112} are
$V=-400\,{\rm cm}^{-1}$ for exciton transfer,
$U=200\, {\rm cm}^{-1}$ for exciton Coulomb interaction,
and $\mu_{2z}/\mu_{1z}=0.35$ for the dimer transition dipoles
orientation asymmetry.
The  characteristic  frequency of system is
$\Omega_s=\sqrt{(\epsilon_1-\epsilon_2)^2+4V^2}=2V$.
The on-site Drude fluctuation parameters are
$\gamma=\gamma_1=\gamma_2=600\,{\rm cm}^{-1}$
and $\lambda=\lambda_1=\lambda_2=600\,{\rm cm}^{-1}$.

 The pump field is a transform-limited Gaussian
pulse with 50\,fs at the full width at half maximum,
centered at $\omega=\epsilon$; i.e.,
$\Delta\omega\equiv \omega-\epsilon=0$,
as indicated by the arrow in panel (a) of each
\Fig{fig4} and \Fig{fig5}. The peak intensity
is of $\mu_{1z}E_{\rm max}=100\,$cm$^{-1}$.
At $t_d=100$\,fs
the pump-transferred occupations
in the single on-site exciton $|1\ra$ and $|2\ra$ states,
and the double-exciton $|f\ra$ state are
6.0\%, 5.9\%, and 1.7\%, respectively, at $T=298$\,K;
while they are 6.3\%, 6.1\%, and 1.9\%, respectively, at
$T=77$\,K.
The probe field assumes in the weak and impulsive limit.
Apparently, the dips appearing in the
nonlinear absorptive $\alpha^{\rm NL}_{\rm A}(\omega,t_d)$
components in the (b) panels
arise mainly from the excite-state absorption to the doubly excited state.
The linear emission signal [dash--curve in (a)] involves
the transition from the lower lying single exciton eigenstate
to ground state, without involving the the double-exciton $|f\ra$ state.
However, as the moderately strong pump field is used,
the nonlinear emissive $\alpha^{\rm NL}_{\rm E}(\omega,t_d)$
component [(c) panels] contains (in the blue side)
also the contribution of the $|f\ra$ state emission.

\begin{figure}
\includegraphics[width=1.0\columnwidth]{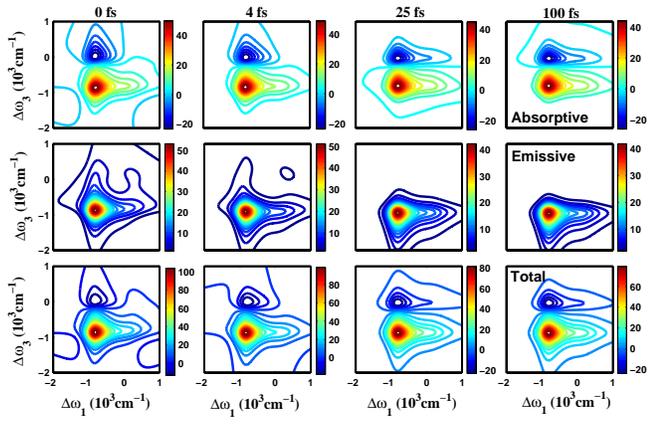}
\caption{Two-dimensional spectroscopy $S(\omega_3,t_d,\omega_1)$
of the same dimer system of \Fig{fig4} at $T=298$\,K.
Both pump and probe fields are weak and in the impulsive
limit.
}
\label{fig6}
\end{figure}

 Figure \ref{fig6} shows the two-dimensional (2D)
spectroscopy $S(\omega_3,t_d,\omega_1)$
of the same dimer system at $T=298$\,K.
As inferred  from \Fig{fig4}, the $[1/1]$ PSD-based
HEOM is sufficient to give the numerically accurate
results here.
Both pump and probe fields are now operated
in the weak and impulsive limit.
As a result,
$S(\omega_3,t_d,\omega_1)$ resolves simply both
the excitation and detection frequencies,
$\omega_1$ and $\omega_3$, of the third-order
optical response function, where $t_2$ is just
the delay time $t_d$ of detection.\cite{Abr092350}
For demonstration  we
examine again the absorptive (upper panels) and
the emissive (middle panels) components of the experimentally
measurable
$S(\omega_3,t_d,\omega_1)$ (bottom panels) that
amounts to the $k_I+k_{II}$ signal.\cite{Abr092350}
Evidently the dips appearing in the absorptive components
are due to the excited state absorption,
while the 2D emissive pathways do not include
the $|f\ra$-state emission, as the pump field
is now in the weak response regime.

\begin{figure}
\includegraphics[width=0.95\columnwidth]{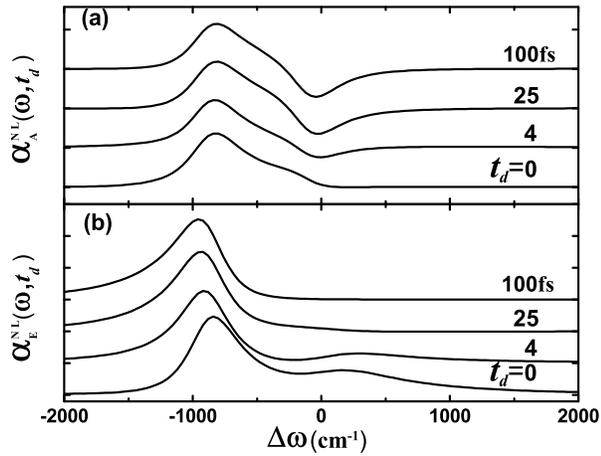}
\caption{The $\Delta\omega_1=0$ slices of
the absorptive and emissive panels of \Fig{fig6}.
This figure resembles the impulsive pump counterpart
of \Fig{fig4}.
}
\label{fig7}
\end{figure}

 Figure \ref{fig7} depicts the $\Delta\omega_1=0$ slices of
the absorptive and emissive panels of \Fig{fig6}.
It thus resembles the impulsive pump counterpart
of \Fig{fig4}, except for the emissive contribution from
the doubly excited state.
Unlike the impulsive limit that considers only sequential
processes,\cite{Muk95} the evaluation with finite pulse duration
involves also coherent processes that cannot be
neglected at the short time
region ($t_d\sim0$ here).\cite{Muk95,Yan906485,Con947855,Kje06024303}

\section{Summary}
\label{thsum}

 In summary, we have constructed the $\NN$ PSD-based HEOM,
which is qualified to be the best hierarchical
theory for Drude dissipation. The present work generalizes
the previous $[0/0]$ and $[1/1]$ PSD-based
hierarchical constructions.\cite{Xu09214111,Tia10114112}
The proposed accuracy control criterions [\Eqs{crit5} and (\ref{crit2})]
are confirmed via not only the reduced system density matrix
dynamics but also nonlinear optical spectroscopy calculations.
No expensive convergency check
would thus be needed for the HEOM dynamics of complex open systems.
HEOM for non-Drude environments that
should be optimized case by case
together with accuracy control criterions
will be considered elsewhere.

\acknowledgments
 Support from the NNSF of China (21033008 \& 21073169),
the National Basic Research Program of China,
(2010CB923300 \& 2011CB921400), and
the Hong Kong RGC (604709) and UGC (AoE/P-04/08-2)
is gratefully acknowledged.


\vspace{1.0 cm}
\newpage
\pagebreak

\end{document}